\documentclass{jaa}
\usepackage{natbib}
\usepackage{multicol}
\usepackage{graphicx}
\usepackage{multirow}
\usepackage{subfigure}
\usepackage{xcolor}
\usepackage{adjustbox}
\usepackage[colorlinks=true,linkcolor=blue,citecolor=blue]{hyperref}
\begin{document}

\title{ACA observation and chemical modelling of phosphorus nitride (PN) towards the hot molecular cores G10.47+0.03 and G31.41+0.31}

\author{Arijit Manna\textsuperscript{1}, Sabyasachi Pal\textsuperscript{1,*}}
\affilOne{\textsuperscript{1}Department of Physics and Astronomy, Midnapore City College, Paschim Medinipur, West Bengal, India 721129\\}

\twocolumn[{
\maketitle
\corres{sabya.pal@gmail.com}


\vspace{0.5cm}
\begin{abstract}
Phosphorus (P) is one of the important elements for the formation of life and plays a crucial role in several biochemical processes. Recent spectral line surveys have confirmed the existence of P-bearing molecules, especially PN and PO, in the star-formation regions, but their formation mechanisms are poorly understood. The P-bearing molecule phosphorus nitride (PN) is detected in several star-forming regions, but this molecule has been poorly studied at high gas densities ($\geq$10$^{6}$ cm$^{-3}$) hot molecular cores. In this article, we present the detection of the rotational emission line of PN with transition J = 3--2 towards the hot molecular cores G10.47+0.03 and G31.41+0.31, using the Atacama Compact Array (ACA). The estimated column densities of PN for G10.47+0.03 and G31.41+0.31 using the local thermodynamic equilibrium (LTE) model are (3.60$\pm$0.2)$\times$10$^{13}$ cm$^{-2}$ and (9.10$\pm$0.1)$\times$10$^{12}$ cm$^{-2}$ with an excitation temperature of 150$\pm$25 K. The fractional abundance of PN relative to H$_{2}$ is 2.76$\times$10$^{-10}$ for G10.47+0.03 and 5.68$\times$10$^{-11}$ for G31.41+0.031. We compute the two-phase warm-up chemical model of PN to understand the chemical evolution in the environment of hot molecular cores. After chemical modelling, we claim that PN is created in the gas phase via the neutral-neutral reaction between PO and N in the warm-up stage. Similarly, PN is destroyed via the ion-neutral reaction between H$_{3}$O$^{+}$ and PN.
\end{abstract}

}]
\doinum{xyz/123}
\artcitid{\#\#\#\#}
\volnum{000}
\year{2021}
\pgrange{1--11}
\setcounter{page}{1}
\lp{11}

	\section{Introduction}
\label{sec:intro}
The identification of new interstellar complex organic molecules gives us an idea of the prebiotic chemistry in star-formation regions, which allows us to learn how the building blocks of life may have evolved in the interstellar medium (ISM) \citep{her09}. Phosphorus (P) is a fundamental element for the study of the origin of life in the universe. Phosphorus is an important compound in the formation of nucleic acids, phospholipids, adenosine triphosphate (ATP), and deoxyribonucleic acid (DNA) \citep{mac97, pas05}. In biotic and prebiotic chemistry, phosphorus gas-phase chemistry is poorly understood because of the minimal observations of its chemical compounds in the ISM. Phosphorus is synthesised in massive stars and injected into the ISM via supernova explosions \citep{ko13, ro14}. Phosphorus has a relatively low abundance relative to hydrogen (2.8$\times$10$^{-7}$), which is lower than that of iron, magnesium, sodium, calcium, and aluminium \citep{gre98}. The detection of phosphorus is difficult because it is heavily depleted in cold and dense molecular clouds by a factor of 600 \citep{tur90, wak08}. In the ISM, only a few P-bearing species, i.e., PN, PO, CP, HCP, C$_{2}$P, and PH$_{3}$ have been detected in the envelopes of evolved stars \citep{ten07, de13, ag14}. Among the detected P-bearing molecules, PN and PO are reported to exist in the star-forming regions (see \citet{riv16} and references therein).

\begin{figure*}
	\centering
	\includegraphics[width=0.95\textwidth]{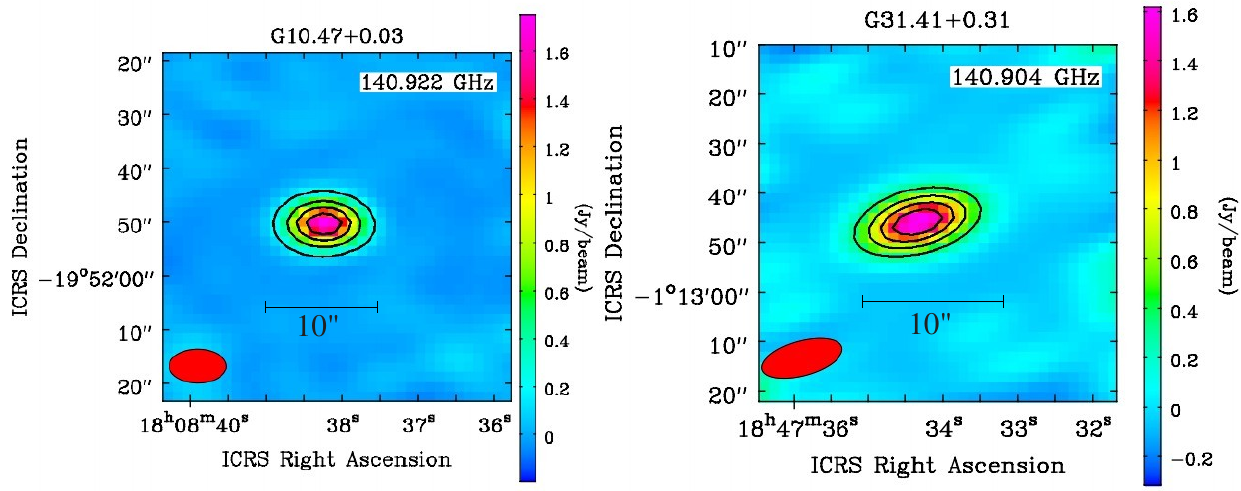}
	\caption{Millimeter wavelength continuum emission images of hot molecular core G10 and G31 are obtained using ALMA ACA band 4 with frequencies of 140.92 GHz and 140.90 GHz. The contour levels start at 3$\sigma$, where $\sigma$ is the RMS of each continuum image. The RMS ($\sigma$) of the continuum emission images of G10 and G31 was 12.27 mJy and 20.98 mJy. Contour levels increased by a factor of $\surd$2. The red circles indicate the synthesized beams of the continuum images.}
	\label{fig:continuum}
\end{figure*}

In ISM, P and N-bearing molecule phosphorus nitride (PN) was first detected in three high-mass star-formation regions, i.e., Orion KL, Sgr B2, and W51, with an estimated abundance of $\sim$(1--4)$\times$10$^{-10}$ \citep{tur87, ziu87, riv16}. The abundance of PN towards Orion KL, Sgr B2, and W51 was larger than the theoretically expected abundance from a pure low-temperature ion-molecule chemical network \citep{tur87, ziu87, riv16}. Previously, the rotational emission lines of PN were identified in several high-mass star-formation regions \citep{tur90, fot16, min18, riv20, ber21}, carbon- and oxygen-rich stars \citep{mil08, de13}, protostellar shocks \citep{lef16}, Galactic Center molecular clouds \citep{riv18}, and class I low-mass protostar B1 \citep{ber19}. Recently, rotational emission lines of PN were also detected in two giant molecular clouds of the sculptor galaxy NGC 253 \citep{has22}. Owing to the limited number of observations of the emission lines of PN and other P-bearing molecules in the ISM, the formation mechanisms of PN and other P-bearing molecules are strongly debated. Previously, three possible formation routes of PN towards the star-formation regions have been proposed: 1. gas-phase chemistry after the shock-induced desorption of P-bearing species (e.g., PH$_{3}$) \citep{aot12, lef16}; 2. high-temperature gas-phase formation routes after the thermal desorption of PH$_{3}$ from ice \citep{ch12}; and 3. gas-phase chemistry of PN and PO during the cold collapse phase and thermal desorption ($\gtrsim$35 K) due to the heat of the protostellar core \citep{riv16}.

In this article, we present the study of the rotational emission line of PN towards the well-known chemically rich hot molecular cores G10.47+0.03 (hereafter G10) and G31.41+0.31 (hereafter G31). The hot molecular cores G10 and G31 are located at a distance of 8.6 kpc \citep{san14} and 7.9 kpc \citep{chu90} from Earth. These hot molecular cores are known to be nurseries of complex organic molecules, including the possible precursors of the simplest amino acid glycine (NH$_{2}$CH$_{2}$COOH) in their warm inner regions \citep{suz16, oh19, man22a}. Earlier, \citet{man22a, man22b}, \cite{man23}, and \citet{min20} discussed the physical and chemical properties of G10 and G31 at millimeter wavelengths. The observations and data reduction are presented in Section~\ref{obs} The result of the detection of rotational emission lines of PN is shown in Section~\ref{res} The discussion and conclusion of the detection of PN towards the G10 and G31 are shown in Section~\ref{dis} and ~\ref{con}
 
\begin{figure*}
	\centering
	\includegraphics[width=0.95\textwidth]{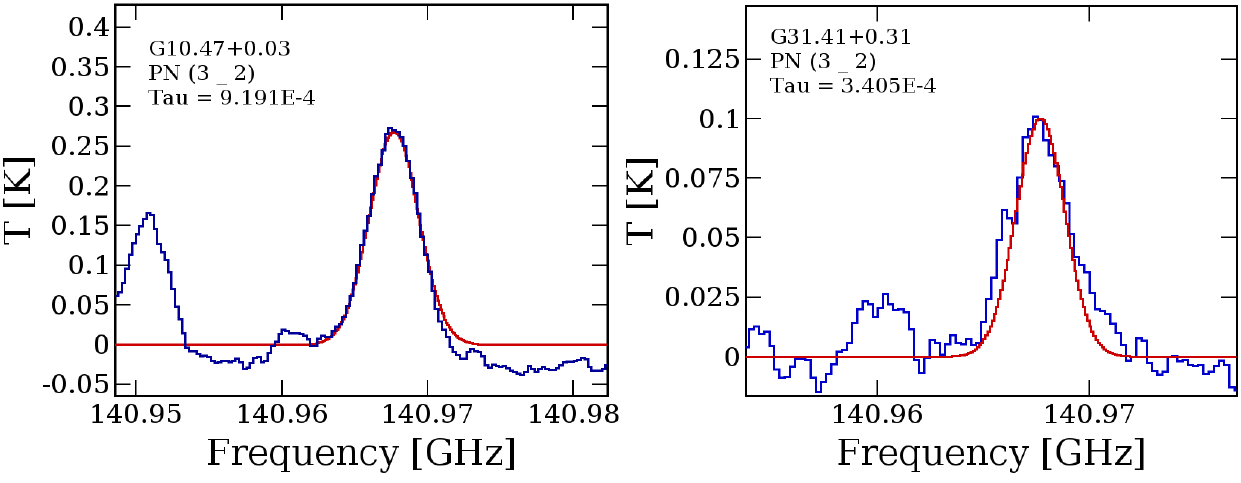}
	\caption{Identified rotational emission line of PN (3--2) towards G10 and G31. The blue spectra indicate the millimeter-wavelength molecular spectra of G10 and G31, and the red spectra represent the best-fit LTE model spectrum for PN. The radial velocity of the G10 and G31 spectra are 68.50 km s$^{-1}$ and 97.0 km s$^{-1}$, respectively.}
	\label{fig:ltespec}
\end{figure*} 

\section{Observations and data reductions}
\label{obs}
We used the archival data of hot molecular cores G10 and G31, which were observed using the Atacama Compact Array (ACA) with 7-metre array antennas (PI: Rivilla, Victor). ACA is the heart of the Atacama Large Millimeter/Submillimeter Array (ALMA). The observation was performed to detect the emission lines of PN with transition J = 3--2 and the spatial distribution of PN towards the G10, G31, and nine other hot molecular cores. We chose only G10 and G31 from that survey data because the luminosities of G10 and G31 are higher than those of other observed sources. The phase center of G10 and G31 were ($\alpha,\delta$)$_{\rm J2000}$ = 18:08:38.232, --19:51:50.400 and 18:47:34.000, --01:12:45.000 respectively. The observation of hot core G10 was carried out on September 16, 2017, using 11 antennas. Similarly, G31 was observed on August 29, 2017, using ten antennas. The observations were made with spectral ranges of 127.47--128.47 GHz, 129.74--130.74 GHz, 139.07--140.07 GHz, and 140.44--141.44 GHz. For the observation of G10, the flux and bandpass calibrator were J1924--2914, and the phase calibrator was J1833--210 B. For the observation of G31, the flux calibrator was taken as Neptune, the bandpass calibrator was taken as J1924--2914, and the phase calibrator was taken as J1851+0035.

For data reduction and spectral imaging, we used the Common Astronomy Software Application (CASA 5.4.1) with an ALMA data reduction pipeline \citep{mc07}. For flux calibration of G10 and G31 using the flux calibrators J1924--2914 and Neptune, we used the Perley–Butler 2017 (for J1924--2914) and Butler-JPL-Horizons 2012 (for Neptune) flux calibrator models for each baseline to scale the continuum flux density of the flux calibrators using the CASA task {\tt SETJY} with 5\% accuracy \citep{per17, but12}. We made the flux and bandpass calibration after the flagging of the bad channels and antenna data using the CASA pipeline with tasks {\tt hifa\_bandpassflag} and {\tt hifa\_flagdata}. We used task {\tt MSTRANSFORM} with all available rest frequencies to separate the target data of G10 and G31 after the initial data reduction. We used the task {\tt UVCONTSUB} in the UV plane of the separated calibrated data of G10 and G31 for continuum subtraction. We created a continuum emission map of G10 and G31 using the CASA task {\tt TCLEAN} with {\tt HOGBOM} deconvolver. To create the spectral data cubes of G10 and G31, we used task {\tt TCLEAN} with cube spectral definition mode {\tt SPECMODE}. Finally, we used the CASA task {\tt IMPBCOR} to correct the primary beam pattern for both the continuum and spectral data cubes.

\section{Result}
\label{res}
\subsection{Continuum emission towards the G10 and G31}
We have shown the millimeter-wavelength continuum emission images of hot core G10 and G31 at frequencies of 140.92 GHz and 140.90 GHz in Figure~\ref{fig:continuum}, where the surface brightness colour scale has the unit of Jy beam$^{-1}$. After creating the continuum emission map of G10 and G31, we fitted the 2D Gaussian over the continuum emission images of G10 and G31 using the CASA task {\tt IMFIT}. We estimate the peak flux density in Jy beam$^{-1}$, integrated flux density in Jy, synthesized beam size in arcsec ($^{\prime\prime}$), deconvolved beam size in arcsec ($^{\prime\prime}$), position angle in degree ($^{\circ}$), and RMS in mJy of the G10 and G31. For G10, the peak flux density and integrated flux density are 1.78$\pm$0.01 Jy beam$^{-1}$ and 1.99$\pm$0.02 Jy with RMS 12.27 mJy and position angle 89.79$^{\circ}$. The deconvolved source size of the continuum emission map of G10 is 2.72$^{\prime\prime}$$\times$2.31$^{\prime\prime}$ while the synthesized beam of the continuum image of G10 is 10.86$^{\prime\prime}$$\times$6.82$^{\prime\prime}$. Similarly, for G31, the peak flux density and integrated flux density are 1.67$\pm$0.02 Jy beam$^{-1}$ and 2.24$\pm$0.04 Jy with RMS 20.98 mJy and position angle --74.44$^{\circ}$. The deconvolved source size of the continuum emission map of G31 is 6.62$^{\prime\prime}$$\times$5.10$^{\prime\prime}$ while the synthesized beam of the continuum image of G31 is 17.89$^{\prime\prime}$$\times$8.77$^{\prime\prime}$. We find that the continuum emission regions of G10 and G31 are smaller than the synthesized beam size, which is estimated after fitting the 2D Gaussian over the continuum emission regions of G10 and G31. It indicates that the continuum emission maps of G10 and G31 are not resolved at frequencies of 140.92 GHz and 140.90 GHz, respectively.

\begin{table}{}
	\centering
	\caption{Observed hyperfine lines of PN.}
	\begin{adjustbox}{width=0.48\textwidth}
		\begin{tabular}{cccccccccccc}
			\hline
			Frequency		& &Transition$^{*}$&&& $E_{u}$ & log$_{10}$($A_{ij}$) &$S\mu^{2}$\\
			
			(MHz)&		($J^{\prime}$&$F^{\prime}$&$J^{\prime\prime}$&$F^{\prime\prime}$) &(K)& &(Debye$^{2}$) \\
			\hline
			140966.005&3&3&2&3&13.530&--5.408&0.8383\\
			140967.684&3&2&2&1&13.530&--4.526&4.5268\\
			140967.684&3&3&2&2&13.530&--3.977&22.635\\
			140967.684&3&4&2&3&13.530&--4.454&9.700\\
			140970.005&3&2&2&2&13.530&--5.262&0.838\\
			
			\hline
		\end{tabular}	
	\end{adjustbox}
	
	
	\label{tab:MOLECULAR DATA}
	{{*}} The J = 3--2 frequency without the hyperfine structure is 140967.684 MHz \citep{caz06}. The hyperfine spectroscopic line parameters of PN are taken from the \href{https://splatalogue.online/advanced1.php}{Splatalogue} and \citet{caz06}.\\
\end{table}		

\subsection{Identification of PN towards the G10 and G31}
At first, we extract the molecular spectra of G10 and G31 from the spectral data cubes to create 23.55$^{\prime\prime}$ and 34.02$^{\prime\prime}$ diameter circular regions over G10 and G31. The synthesized beam sizes of the spectral data cubes of G10 and G31 are 10.78$^{\prime\prime}$$\times$6.28$^{\prime\prime}$ and 17.03$^{\prime\prime}$$\times$7.06$^{\prime\prime}$. The systematic velocity ($V_{LSR}$) of the G10 and G31 is 68.50 km s$^{-1}$ and 97.0 km s$^{-1}$ \citep{rof11, suz16}. To detect the rotational emission line of PN in the spectra of G10 and G31, we used the local thermodynamic equilibrium (LTE) model spectra with the Cologne Database for Molecular Spectroscopy (CDMS) \citep{mu05}. We used the CASSIS software for LTE spectral modelling of PN \citep{vas15}. The LTE assumptions are valid in the inner regions of G10 and G31 because the gas density of the warm inner regions of G10 and G31 is $\sim$1.2$\times$10$^{^7}$ cm$^{-3}$ \citep{rof11, suz16, min20}. After LTE spectral analysis, we identify the rotational emission line of PN (J = 3--2) and its hyperfine transitions in the spectra of G10 and G31. For fitting the LTE model over the observed spectra of PN, we use Markov chain Monte Carlo (MCMC) in CASSIS. The MCMC technique uses a random walk to iteratively go over all of the parameters and move into the solution space, making it the ultimate outcome by $\chi^{2}$ minimization. The MCMC method considers choosing an initial value in a five-dimensional parameter space, which we call the X$_{0}$ state. Subsequently, it chooses an arbitrary closest neighbour, which is referred to as the X$_{1}$ state. This state is based on a variable step size that is estimated for each iteration. Using this method, $\chi^{2}$ of the new state is estimated, and if $p = \chi^{2}$(X$_{0}$)/$\chi^{2}$(X$_{1}$) $>$1 then the new state is accepted. If $p = \chi^{2}$(X$_{0}$)/$\chi^{2}$(X$_{1}$) $<$ 1 then the new state may be accepted with a certain acceptance probability. If the new state is rejected, the existing X$_{0}$ state will remain, and the X$_{1}$ state will be chosen randomly from among the neighbouring states. This algorithm runs with several initial random states, and it is typically deemed to have reached the correct solution when the variation between different groups of states is less than the variance of each group. All spectral parameters, such as excitation temperature, column density, source size, FWHM, and V$_{LSR}$, can be varied. The partition functions are computed using CASSIS for temperatures lower than 9.375 K, which is the lowest temperature provided by the CDMS database for some molecules. The partition function
can be computed as:
	
\begin{equation}
Q(T) = \sum g_{i}\times exp(-E_{i}/kT)
\end{equation}
where $E_{i}$ and $g_{i}$ are the energy and degeneracy of level $i$, respectively.

\begin{figure*}
	\centering
	\includegraphics[width=1.0\textwidth]{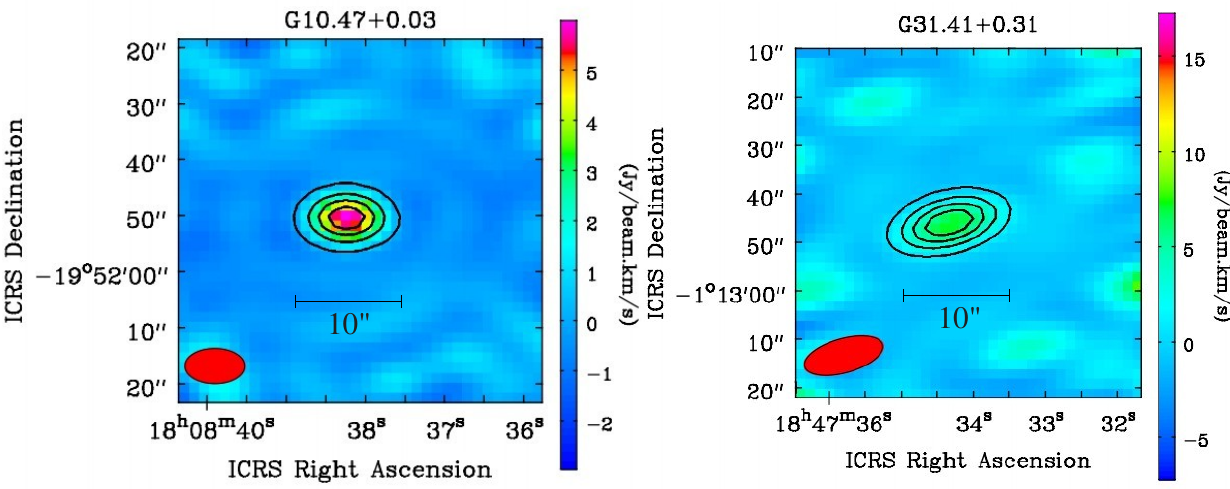}
	\caption{Integrated emission map of PN with transition J = 3--2 towards the G10 and G31, which was overlaid with the 140.92 GHz and 140.90 GHz continuum emission map of G10 and G31. The contour levels were 20\%, 40\%, 60\%, and 80\% of peak flux. The red circles represent the synthesized beam of the integrated emission maps. The peak flux density of the emission map of PN towards G10 and G31 was 6.12$\pm$0.39 Jy beam$^{-1}$ km s$^{-1}$ and 8.66$\pm$0.24 Jy beam$^{-1}$ km s$^{-1}$}.
	\label{fig:emissionmap}
\end{figure*}
	
The J = 3--2 transition line of PN splits into five hyperfine components ($F^{\prime}$ and $F^{\prime\prime}$) owing to the nuclear quadrupole interaction of the N nucleus ($I$ = 1), as shown in Table~\ref{tab:MOLECULAR DATA}. Here, we use the rest frequency of the PN (J = 3--2) line without the hyperfine structure at a frequency of 140967.684 MHz taken from the CDMS and \citet{caz06}. After the LTE analysis, we find that the hyperfine transitions of PN towards G10 and G31 do not resolve due to insufficient spectral resolution to resolve the hyperfine lines of PN. Additionally, we also check for the possible blending effect in the PN emission spectra with nearby molecular transitions using CDMS databases, but we find that the identified emission spectra of PN do not blend with nearby molecular lines. The full-width half maximum (FWHM) of the LTE-fitted rotational emission spectra of PN towards the G10 and G31 are 8.50 km s$^{-1}$ and 5.80 km s$^{-1}$, respectively. The best-fit column density of PN towards G10, using the LTE modelled spectra, is (3.60$\pm$0.2)$\times$10$^{13}$ cm$^{-2}$ with an excitation temperature of 150$\pm$25 K and a source size of 10.78$^{\prime\prime}$.  Similarly, for G31, the best-fit column density is (9.10$\pm$0.1)$\times$10$^{12}$ cm$^{-2}$ with an excitation temperature of 150$\pm$25 K. Our estimated excitation temperature indicates that the detected emission line of PN originates from the inner region of the hot core because the temperature of the hot molecular core is above 100 K \citep{her09, man23a}. During the estimation of the column density, we assume an excitation temperature of 150$\pm$25 K to estimate the column density of the single transition line of PN. We use the same excitation temperature for the two hot cores because the gas densities of G10 and G31 are nearly similar \citep{suz16}. The LTE-fitted rotational emission spectra of PN towards G10 and G31 are shown in Figure~\ref{fig:ltespec}. The estimated fractional abundances of PN towards G10 and G31 with respect to H$_{2}$ are 2.76$\times$10$^{-10}$ and 5.68$\times$10$^{-11}$ respectively, whereas the column densities of H$_{2}$ towards G10 and G31 are 1.3$\times$ 10$^{23}$ cm$^{-2}$ and 1.6$\times$ 10$^{23}$ cm$^{-2}$, respectively, \citep{ik01}. Earlier, \citet{fot19} first detected the emission line of PN towards G10 and G31 using IRAM with lower excitation temperatures of 16 K and 10 K. The low excitation temperature of PN estimated by \citet{fot19} does not truly represent hot molecular cores because the temperatures of the hot cores are above 100 K \citep{her09}. \cite{fot19} also does not create the emission map of PN due to the restriction of the IRAM resolution. Our ACA observation of PN indicates that the emission line of PN is emitted from the warm-inner regions of G10 and G31.

\subsection{Spatial distribution of PN towards the G10 and G31}
After detecting the rotational emission line of PN, we produce an integrated emission map of PN with transition J = 3--2 using the CASA task {\tt IMMOMENTS} towards G10 and G31. The integrated emission maps of PN towards G10 and G31 are produced by integrating the spectral data cubes in the velocity range in which the emission lines of PN are detected. The integrated emission maps of PN for G10 and G31 are shown in Figure~\ref{fig:emissionmap}. The velocity ranges of the integrated emission maps of PN towards G10 and G31 are 60 to 75 km s$^{-1}$ and 90 to 105 km s$^{-1}$, respectively. The resultant integrated emission maps of PN towards G10 and G31 are overlaid with the 140.92 GHz and 140.90 GHz continuum emission maps of G10 and G31. We observe that the integrated emission maps of PN towards G10 and G31 had a peak at the continuum position. From the integrated emission maps, it is evident that PN originate from the warm inner part of the hot core regions of G10 and G31. After producing the integrated emission maps of PN, we estimate the emitting regions of PN towards G10 and G31 by fitting the 2D Gaussian over the integrated emission maps of PN using the CASA task {\tt IMFIT}. The deconvolved beam size of the emitting region of PN is estimated by the following equation,
	\begin{equation}
	\theta_{S}=\sqrt{\theta^2_{50}-\theta^2_{beam}}
	\end{equation}
where $\theta_{50} = 2\sqrt{A/\pi}$ indicates the diameter of the circle whose area ($A$) encloses the 5\% line peak of PN, and $\theta_{beam}$ is the half-power width of the synthesized beam \citep{man22a}. The estimated emitting regions of PN towards the G10 and G31 are 0.44 pc and 0.65 pc, respectively. We noticed that the emitting regions of PN towards G10 and G31 are similar or small with respect to the synthesized beam size. This indicates that the J = 3--2 transition line of PN is not spatially resolved or, at best, marginally resolved. Therefore, we cannot draw any conclusions about the morphology of the integrated emission maps of PN towards G10 and G31. The higher spatial and spectral resolution observation of PN using the ALMA 12-m array is needed to draw any conclusions about the chemical morphology of PN towards G10 and G31.
	
\begin{figure*}
	\centering
	\includegraphics[width=1.0\textwidth]{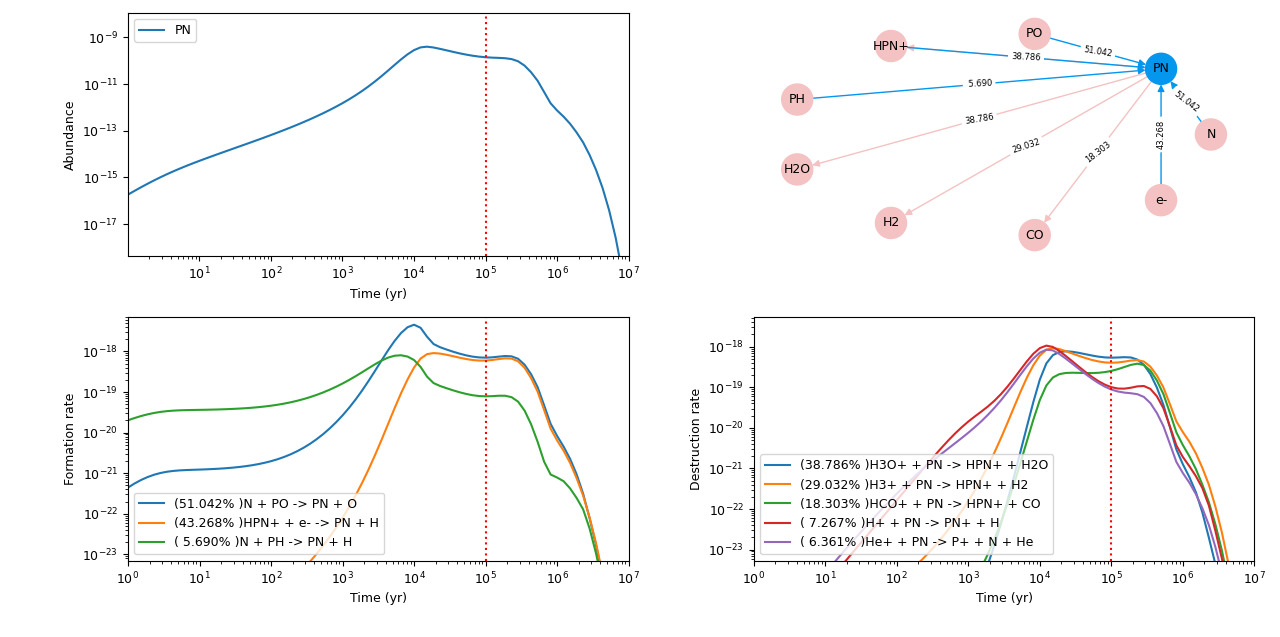}
	\caption{Two-phase warm-up chemical evolution of the abundance of PN obtained with the GGCHEMPY chemical code for hot molecular cores. The destruction pathways of PN are shown in the lower-right panel. The abundance of PN varies between 8.20$\times$10$^{-12}$ to 3.90$\times$10$^{-10}$ for such conditions.}
	\label{fig:chemical model}
\end{figure*}

\section{Discussion}
\label{dis}
\subsection{Previous chemical modelling of PN in the high-mass star-forming regions}
Previously, \citet{riv16} and \citet{jim18} computed the two-phase warm-up chemical model of PN using the gas-grain chemical kinetics code UCLCHEM \citep{hol17} in the environment of high-mass star-formation regions. During chemical modelling, \citet{riv16} and \cite{jim18} assumed the formation of a dark, dense clump from the translucent cloud during free-fall collapse. In the first phase of chemical modelling, the gas density increased from 10$^{2}$ to 10$^{6}$ cm$^{-3}$ and the visual extinction (A$_{v}$) increased from 2 to 100 mag. Similarly, the dust and gas temperatures decreased from 20 to 10 K. In the second phase, the dense dark clump warms up to 10 K to 200 K during 2$\times$10$^{5}$ yr. The second phase is known as the warm-up protostellar phase, where the hot core was developed during 2$\times$10$^{5}$ yr. After chemical modelling, \citet{riv16} and \cite{jim18} observed that the abundance of PN varies from 10$^{-12}$ to almost 10$^{-9}$. To estimate the modelled abundance of PN, \citet{riv16} and \cite{jim18} used the gas-phase neutral-neutral reaction between PO and N at 1.2$\times$10$^{5}$ yr (warm-up phase). \citet{riv16} and \cite{jim18} claim that the abundances of the PO and PN become almost equal until 1.2$\times$10$^{5}$ yr. In the warm-up phase, the gas temperature gradually increased from 10 to 200 K, and the species evaporated to the gas phase, while PN and PO were desorbed when the temperature reached $\sim$35 K. The possible gas-phase reactions are as follows.
	\\\\
PO + N $\rightarrow$ PN + O ~~~~~~~~~~~~~~~(1)\\\\
Recently, \citet{sil21} claimed that reaction 1 is the most efficient pathway for the formation of PN in high-mass star-formation regions and hot molecular cores. Similarly, \citet{riv16} claimed that PO can be efficiently created during the cold collapse phase in the star-formation regions in a sequence of gas-phase ion-molecule reactions.\\\\
H$_{3}$O$^{+}$ + P $\rightarrow$ HPO$^{+}$ + H$_{2}$~~~~~~(2)\\\\
HPO$^{+}$ + e$^{-}$ $\rightarrow$ PO + H ~~~~~~~~~~(3)\\\\
and\\\\
P$^{+}$ + H$_{2}$ $\rightarrow$ PH$_{2}$$^{+}$~~~~~~~~~~~~~~~~~~(4)\\\\
PH$_{2}$$^{+}$ + e$^{-}$ $\rightarrow$ PH + H~~~~~~~~~~(5)\\\\
PH + O $\rightarrow$ PO + H~~~~~~~~~~~~~~~(6)\\\\
Similarly, \citet{riv16} claimed that the PN would be destroyed when protonated water (H$_{3}$O$^{+}$) reacts with the PN in the gas phase via the following reaction\\\\
H$_{3}$O$^{+}$ + PN $\rightarrow$ HPN$^{+}$ + H$_{2}$O~~~~~~~~~~~~~~~(7)\\\\
where after the dissociative recombination of HPN$^{+}$, two equally probable reactions were created \citep{riv16}.\\\\
HPN$^{+}$ + e$^{-}$ $\rightarrow$ PN + H ~~~~~~~~~~(8)\\\\
and\\\\
HPN$^{+}$ + e$^{-}$ $\rightarrow$ PH + N ~~~~~~~~~~(9)\\\\
Since the fractional abundance of oxygen is higher than the abundance of nitrogen, PH is converted to PO instead of PN. Therefore, PN is gradually destroyed, but PO is produced significantly, which increases the PO/PN ratio in the star-formation regions (see Figure 7(d) in \citet{riv16}).

\begin{table*}{}
	\centering
	\scriptsize
	\caption{Formation and destruction pathways of PN with rate coefficients.}
	\begin{adjustbox}{width=0.95\textwidth}
		\begin{tabular}{|c|c|c|c|c|c|c|c|c|c|c|c|}
			\hline 
			&Reaction&Types&$\alpha$&$\beta$&$\gamma$\\
			\hline
			Formation&N + PO $\rightarrow$ PN + O&Neutral-neutral reaction between N and PO&3.01$\times$10$^{-11}$&-0.60&0.00\\
			&HPN$^{+}$ + e$^{-}$ $\rightarrow$ PN + H&Dissociative recombination of HPN$^{+}$&1.00$\times$10$^{-7}$&-0.50&0.00\\	
			&N + PH $\rightarrow$ PN + H&Neutral-neutral reaction between N and PH&5.00$\times$10$^{-11}$&0.00&0.00\\	
			\hline
			Destruction&H$_{3}$O$^{+}$ + PN $\rightarrow$ HPN+ + H$_{2}$O&Ion-neutral reaction between H$_{3}$O$^{+}$ and PN&1.00$\times$10$^{-9}$&-0.50&0.00\\
			&H$_{3}$$^{+}$ + PN $\rightarrow$ HPN+ + H$_{2}$&Ion-neutral reaction between H$_{3}$$^{+}$ and PN&1.00$\times$10$^{-9}$&-0.50&0.00\\
			&HCO$^{+}$ + PN $\rightarrow$ HPN+ + CO&Ion-neutral reaction between HCO$^{+}$ and PN&1.00$\times$10$^{-9}$&-0.50&0.00\\
			&H$^{+}$ + PN $\rightarrow$ PN$^{+}$ + H&Charge exchange of PN&1.00$\times$10$^{-9}$&-0.50&0.00\\
			&He$^{+}$ + PN $\rightarrow$ P$^{+}$ + N + He &Ion-neutral reaction between He$^{+}$ and PN&1.00$\times$10$^{-9}$&-0.50&0.00\\
			\hline
		\end{tabular}
	\end{adjustbox}	
	
	\label{tab:reactions}
\end{table*}
	
The chemical modelling of \citet{riv16} and \cite{jim18} is not valid for G10 and G31 because the gas densities of G10 and G31 are $\sim$1$\times$10$^{7}$ cm$^{-3}$ and during the warm up stage \citet{riv16} and \cite{jim18} increase the gas density upto 10$^{6}$ cm$^{-3}$. To understand the formation mechanism of PN towards G10 and G31, new chemical modelling is needed with the gas density of 1$\times$10$^{7}$ cm$^{-3}$ and gas temperature 300 K (typical hot core temperature).

\subsection{New chemical modelling of PN towards hot molecular cores}
We made the two-phase warm-up chemical modelling of PN to estimate the modelled abundance from the reactions and uncover the proper formation mechanism of PN towards the hot molecular cores.  We use the gas-grain chemical code GGCHEMPY for chemical modelling \citep{ge22}. During warm-up chemical modelling, the free-fall collapse of the cloud (Phase I) is followed by a warm-up phase (Phase II). Our two-phase warm-up chemical model is similar to that of \cite{vi04}. In phase I, the gas density rapidly increase from $n_{H}$ = 300 cm$^{-3}$ to 1$\times$10$^{7}$ cm$^{-3}$ and the dust temperature remained constant at 10 K. During modelling, the initial visual extinction ($A_{V}$) is assumed to be 2, and the cosmic ray ionization rate is assumed to be 1.3$\times$10$^{-17}$ s$^{-1}$ \citep{vi04}. During this time, atoms and molecules are accreted on the grain surface of the hot molecular cores. The accretion rate depends on the gas density of the hot cores. At this stage, the chemical species may hydrogenate or rapidly react with other compounds on the grain surface. In our chemical model, the abundances of oxygen (O), carbon (C), nitrogen (N), and helium (He) corresponding to the solar values are obtained from \cite{as09}. Other atomic compounds, such as sulphur (S), silicon (Si), magnesium (Mg), chlorine (Cl), fluorine (F), and phosphorus (P) are depleted by factors of 100. In phase II, the gas density remains constant at 1$\times$10$^{7}$ cm$^{-3}$ and the gas temperature increases from 10 to 300 K during 1$\times$10$^{5}$ yr,  which develops into a hot molecular core (``warm-up protostellar phase"). The following trends are discussed in \cite{vi04}. In phase II, the molecules no longer freeze, and the frozen molecules on the grains are removed in the gas phase by both non-thermal and thermal desorption mechanisms. In our chemical model, we use thermal evaporation analysis as detailed in \cite{vi04}. In our chemical model, we also include co-desorption with H$_{2}$O, volcano desorption, and mono-molecular desorption, which are described in detail in \cite{col04}.

During chemical modelling, we add three formation pathways for the production of PN in the chemical network of GGCHEMPY. Similarly, we add five destruction pathways to the PN. The chemical reactions and reaction coefficients which obtained by the Arrhenius equation are presented in Table~\ref{tab:reactions}. The formation and destruction pathways and the reaction coefficients are obtained from the UMIST 2012 astrochemistry chemical network \citep{mce13}. The computed chemical model is shown in Figure. \ref{fig:chemical model}. After chemical modelling, the modelled abundance of PN in the hot molecular cores varies from 8.20$\times$10$^{-12}$ to 3.90$\times$10$^{-10}$ during the warm-up phase. In our modelling, we see that the neutral-neutral reaction between N and PO produces a sufficient amount of PN in the gas phase of the hot molecular cores. The other two reactions, the dissociative recombination of HPN$^{+}$ and neutral-neutral reactions between N and PH, do not produce a sufficient amount of PN. We also compute the formation and destruction rates with respect to time, corresponding to three formation pathways and five destruction pathways of PN. We notice that the neutral-neutral reaction between N and PO in the gas phase has a 51\% ability to produce PN in hot molecular cores. Similarly, we confirm that the ion-neutral reaction between H$_{3}$O$^{+}$ and PN is the most likely pathway for the destruction of PN.

\subsection{Comparision between observed and modelled abundance of PN}
We compared our estimated abundance of PN towards G10 and G31 with our computed two-phase warm-up chemical model abundance of PN. Our two-phase warm-up chemical model of PN is suitable with G10 and G31 because the gas densities of both hot cores are $\sim$1$\times$10$^{7}$ cm$^{-3}$ and the temperatures of both sources are above 150 K \citep{rof11, min20, man22a}. We estimate that the fractional abundances of PN towards G10 and G31 are 2.76$\times$10$^{-10}$ and 5.68$\times$10$^{-11}$. From the two-phase warm-up chemical modelling, we see that the modelled abundance of PN varies between 8.20$\times$10$^{-12}$ to 3.90$\times$10$^{-10}$ in the warm-up phase. This indicates an estimated abundance of PN towards G10 and G31, which is nearly similar to the range of the modelled abundance of PN.  This result clearly indicates that PN is created towards G10 and G31 via the neutral-neutral reaction between PO and N in the gas phase. Earlier, \cite{tur87} detected the rotational emission lines of PN from Orion (KL), W51-M, and Sgr B2 with fractional abundances of 1.7$\times$10$^{-10}$, 1.1$\times$10$^{-11}$, and 1.7$\times$10$^{-12}$. The abundance of PN towards Orion (KL), W51-M, and Sgr B2 is similar to our range of modelled abundances of PN. That indicates that except for G10 and G31, the P and N-bearing molecule PN is created via the neutral-neutral reaction between PO and N in the gas phase towards other hot cores, Orion (KL), W51-M, and Sgr B2.

\section{Summary and conclusion}
\label{con}
In this article, we present the detection of the rotational emission line of PN with transition J = 3--2 towards the hot molecular cores G10 and G31 using ACA. The estimated column densities of PN towards G10 and G31.41 are (3.60$\pm$0.2)$\times$10$^{13}$ cm$^{-2}$ and (9.10$\pm$0.1)$\times$10$^{12}$ cm$^{-2}$, respectively, with an excitation temperature of 150$\pm$25 K. The abundance of PN towards G10 is 2.76$\times$10$^{-10}$. Similarly, the abundance of PN in G31 is 5.68$\times$10$^{-11}$. We create the integrated emission map of PN towards G10.47 and G31.41, and we observe that the PN arises from the warm inner regions of both hot cores with emitting regions of 0.44 pc and 0.65 pc respectively. We also compute a two-phase warm-up chemical model to understand the formation mechanism of PN. We derive the modelled abundance of PN from chemical modelling. After chemical modelling, we compared our estimated abundance of PN with the modelled abundance of PN and observed that the observed and modelled abundances are nearly similar. After this comparison, we claim that PN is created in the gas phase via the neutral-neutral reaction between PO and N in the warm-up stage of G10 and G31. Except for G10 and G31, we also observe that our modelled abundance is similar to the observed abundance of PN towards Orion (KL), W51-M, and Sgr B2. This result also concludes that the PN is formed towards Orion (KL), W51-M, and Sgr B2 via the neutral-neutral reaction between PO and N. We also search for emission lines of PO towards G10 and G31 to estimate the PO/PN ratio, which will be carried out in our follow-up study.

\section*{Acknowledgement}
We thank the anonymous referee for the helpful comments that improved the manuscript. A.M. acknowledges the Swami Vivekananda Merit cum Means Scholarship (SVMCM), Government of West Bengal, India, for financial support for this research. The plots within this paper and other findings of this study are available from the corresponding author upon reasonable request. This paper makes use of the following ALMA data: ADS /JAO.ALMA\#2016.2.00005.S. ALMA is a partnership of ESO, NSF (USA), and NINS (Japan), together with NRC (Canada), MOST and ASIAA (Taiwan), and KASI (Republic of Korea), in cooperation with the Republic of Chile. The Joint ALMA Observatory is operated by ESO, AUI/NRAO, and NAOJ.

\bibliographystyle{aasjournal}

\end{document}